# Globular Clusters and Satellite Galaxies: Companions to the Milky Way

*Use an interactive model of the Milky Way to discover relationships between our galaxy and its companions.*

by Duncan A. Forbes, Pavel Kroupa, Manuel Metz, and Lee Spitler

Our Milky Way galaxy is host to a number of companions. These companions are gravitationally bound to the Milky Way and are stellar systems in their own right. They include a population of some 30 dwarf satellite galaxies (DSGs) and about 150 globular clusters (GCs).

## Our Galaxy's Companions

Globular clusters — dense, spherical collections from about 10,000 to as many as one million stars — are well known to amateur astronomers. Satellite galaxies include the Large and Small Magellanic Clouds, visible to the naked eye in the Southern Hemisphere, and a host of smaller, fainter galaxies. These DSGs are more extended and diffuse compared to GCs. In fact the larger satellites of the Milky Way (and other galaxies, for that matter) host their own small systems of GCs.

The number of known DSGs and GCs has increased in recent years as a result of new surveys like the Sloan Digital Sky Survey [www.sdss.org] (SDSS) and 2MASS [www.ipac.caltech.edu/2mass]. The SDSS, which has to date surveyed some 20% of the sky at optical wavelengths, is well suited to finding low surface brightness objects. The 2MASS all-sky survey, which operates at infrared wavelengths, has been used to discover several dust-obscured objects near the galactic center. Many more DSGs and GCs likely await detection.

The distinction between DSGs and GCs is being blurred as the Sloan Survey is finding some objects with intermediate luminosities and sizes. These objects share some properties with the so-called Palomar class of extended globular clusters. One of the least luminous known DSGs is called Segue 1 with a V-band luminosity of M = -1.5 (similar to the luminosity of a single supergiant star!).

But could Segue 1 actually be a new globular cluster? There isn't a universally agreed-upon definition for a GC or DSG. However one major difference, and perhaps a defining feature of a galaxy, is the presence of a massive dark-matter halo. Assuming Segue 1 is a stable, bound object, then it appears to have much more mass than light and is therefore a galaxy (albeit a very faint one).

To further complicate matters, the most luminous GC (Omega Centauri [www.spacetelescope.org/news/html/heic0809.html]) is now thought to be the remnant nucleus of a dwarf satellite galaxy that formed billions of years ago but was subsequently gravitationally stripped of its outer stars. Another example of this could be the globular M54, which may once have been the nucleus of the Sagittarius Dwarf Satellite Galaxy (Sgr DSG) that came together in the last few billion years. One clue supporting this interpretation is the presence of multiple stellar populations within these two 'globular clusters.' In the case of the Sgr DSG, a half dozen globulars have been identified as having once belonged to this little galaxy. Thus the Milky Way now plays host to a population of accreted globulars, along with those that were formed in situ. Also, researchers have identified several coherent great circles, or 'streams' of GCs across the sky, which may signpost the past orbits of long-since disrupted dwarf galaxies.

**Making a Dwarf Galaxy**
Two scenarios have been suggested to explain the presence of these dwarf satellite galaxies. The first is that they formed within halos of cold dark matter (CDM) shortly after the Big Bang. Observations of the motions of stars within DSGs are consistent with each dwarf being surrounded by a dark-matter halo — as expected in the CDM scenario. Another prediction from CDM is that while hundreds of satellite dark-matter halos form, only a fraction will contain a dwarf galaxy at their centre. Since the number of DSGs observed is much smaller than the total number of halos predicted, the bulk of the halos must remain dark.

The second scenario is that the Milky Way was involved in an encounter with another large galaxy, and that many of today's dwarf satellites are the tidal debris of this long-ago meeting. Observations [www.spacetelescope.org/images/html/heic0810ab.html] of nearby interacting galaxies show that the tidal debris of star-forming gas is a common occurrence, and in many cases this leads to the formation of so-called tidal dwarf galaxies. Such galaxies are expected to be free of dark matter. In this case the internal motions of stars in DSGs are due to the ongoing gravitational interaction with the Milky Way.

Recently, it has been noticed that many of the Milky Way's dwarf satellite galaxies lie on a single great plane in the sky. If not simply a projection effect or an unlikely random occurrence, then this alignment implies that these galaxies have a similar origin. In the case of an ancient encounter, we would expect the tidal debris to form a great plane in the sky rotating around the Milky Way. The cold dark matter scenario can also produce a great plane of DSGs if the Milky Way captured them as a small coherent group of galaxies. But it does require that the Milky Way had very few original, satellite dark-matter halos that eventually formed galaxies.

For both cases it is interesting to examine the possibility of a spatial connection between the DSGs and GCs of the Milky Way. So lets do that. Exploring a spatial connection between objects in a volume of space such as the halo of the Milky Way is best done in three dimensions. First we need a three-dimensional (3-D) model of the Milky Way and its companions. Fortunately, as in the cooking shows on television, one has already been prepared.

**A 3-D Model of Our Galaxy and its Companions**
Using a new 3-D plotting package called S2PLOT [http://astronomy.swin.edu.au/s2plot], a representation of the Milky Way galaxy and its companions has been created. In the model, the globular clusters appear as small blue and red spheres depending their chemical properties: blue if they are poor in metals (elements heavier than helium) and red if they are metal rich. Dwarf satellite galaxies (and candidates for new DSGs) are assigned larger white spheres. Both GCs and DSGs are placed at their true position in 3-D space based on their measured coordinates in the sky and their estimated distances.

GC data are taken from the compilation of Bill Harris [http://www.physics.mcmaster.ca/~harris/mwgc.dat]. The DSG positions are derived from a variety of sources, with most of the new ones coming from the SDSS survey and 14 DSG candidates kindly supplied by Helmut Jerjen. The model includes a representation of the SDSS survey volume shown as a 'spider web' of green lines. The Milky Way galaxy is represented by a flat image of a spiral galaxy. The position of the Sun is shown by a yellow dot. The size of these objects is not to scale!

The image above shows the Milky Way, and the 3-D distribution of its GCs and DSGs. Click on this hotlink [http://astronomy.swin.edu.au/~dforbes/indexdf.html]

(or the one in the caption) to open up the graphic in a new window. With most Web browsers and Flash 9 [www.adobe.com/products/flashplayer] installed, you should be able to interact with the figure, rotating and zooming to your heart's content.

In the opening view you can see that the GCs seem concentrated towards the Milky Way. While some are fairly remote, it's the DSGs that are spread out to large distances. By interacting with the figure, you can see that they are not uniformly distributed; most lie in a great plane on the sky. More are located in the northern sky as that is the region (shown by the green spider web) probed by the SDSS survey. The known and candidate DSGs show a north-south (minor axis) alignment that passes through the center of Milky Way. This is not simply a selection effect due to the fact that only 20% of the sky has been carefully searched. A thorough examination of the SDSS search region reveals that large areas of the search volume are devoid of galaxies. Thus the plane of these DSGs appears to be a real feature. Furthermore, most of the outer globular clusters also appear to lie in this same great plane. This suggests that these GCs were once associated with the DSGs and have been accreted by the Milky Way, along with the DSGs that they originally belonged to.

**Understanding DSGs**

The nature and origin of the DSGs also has implications for our theories of gravity. If, for example, astronomers conclude that the DSGs are the result of tidal debris, and at the same time, contain much dark matter, then this would strongly indicate that these galaxies are not governed by normal Newtonian gravity. This is because tidal dwarf galaxies cannot contain much dark matter due to their formation from relatively dark-matter-free gas clouds.

The study of satellite galaxies and globular clusters provides an important test of our current theories of cosmology and even gravity. We expect some interesting surprises in the near future as our understanding of these enigmatic objects improves.

*The authors — Duncan Forbes and Lee Spitler (Swinburne University, Australia), Pavel Kroupa and Manuel Metz (Bonn University, Germany) — are interested in both globular clusters and dwarf satellite galaxies and believe they hold important clues to the formation history of our home — the Milky Way galaxy. We would like to thank Andrew Jackling and David Barnes for their input toward the 3-D figure, and Helmut Jerjen for sending us the locations of his dwarf galaxy candidates. A video podcast is available via itunes (www.apple.com/itunes), simply type `Companions to the Milky Way' into the itunes search.*